\documentclass{vgtc}                          

\ifpdf
  \pdfoutput=1\relax                   
  \usepackage{graphicx}                
  \DeclareGraphicsExtensions{.pdf,.png,.jpg,.jpeg} 
\else
  \usepackage{graphicx}                
  \DeclareGraphicsExtensions{.eps}     
\fi%

\graphicspath{{figures/}{pictures/}{images/}{./}} 

\usepackage{microtype}                 
\PassOptionsToPackage{warn}{textcomp}  
\usepackage{textcomp}                  
\usepackage{mathptmx}                  
\usepackage{times}                     
\usepackage{cite}                      
\usepackage{tabu}                      
\usepackage{booktabs}                  
\newcommand\blfootnote[1]{%
  \begingroup
  \renewcommand\thefootnote{}\footnote{#1}%
  \addtocounter{footnote}{-1}%
  \endgroup
}

\onlineid{0}

\vgtccategory{Research}

\vgtcinsertpkg

\title{Design and Development of PainBit: a Portable Device for Supporting Patients with Chronic Pain to Log their Pain}

\author{Arsh Saleem\\ %
        \parbox{1.4in}{\scriptsize \centering Department of Systems and Computer Engineering \\ Carleton University}
 \and Beck Langstone\\ %
         \parbox{1.4in}{\scriptsize \centering Department of Human-Computer Interaction \\ Carleton University}
 \and Alicia Ouskine\\ %
         \parbox{1.4in}{\scriptsize \centering Department of Human-Computer Interaction \\ Carleton University}
 \and Fateme Rajabiyazdi\thanks{e-mail: fateme.rajabiyazdi@carleton.ca}\\ %
         \parbox{1.4in}{\scriptsize \centering Department of Systems and Computer Engineering \\Carleton University}}

\abstract{Recently, we have seen growing interest among patients with
chronic conditions to track their health-related data. There are many wearable devices available to track different health data. However, tracking pain is mostly done by using pen and paper or mobile apps. 
In collaboration with a healthcare professional we designed a portable pain tracker, PainBit. To gain an understanding of patients' perspectives on our tracker, we conducted two case studies with patients living with chronic pain. We asked patients to use PainBit for two weeks and later conducted semi-structured interviews with them. Patients found PainBit useful for tracking their pain and they preferred using a physical device, PainBit, to track their pain over using a mobile phone. Patients suggested reducing the size and weight of PainBit in the next iterations. We report on the lessons learnt through our design process and the evaluation studies.%
} 

\CCScatlist{
  \CCScatTwelve{Human-centered computing}{Visu\-al\-iza\-tion}{Visu\-al\-iza\-tion techniques}{Treemaps};
  \CCScatTwelve{Human-centered computing}{Visu\-al\-iza\-tion}{Visualization design and evaluation methods}{}
}

\begin{document}


\firstsection{Introduction}
\maketitle
Pain is a difficult metric to measure and properly quantify. The measurement of pain is important for chronic pain patients, as tracking pain can give patients a better understanding of how to manage their chronic pain. Additionally, collecting pain tracking data helps patients become active contributors to their own care, and allows healthcare professionals to better support patients, leading to improved quality of care.

Many people develop chronic pain after an injury, illness, or surgery, meaning the pain persists for months after the cause is resolved. Chronic pain syndrome is a condition that keeps many patients in pain throughout their days. These patients can experience pain spikes at any time, and related research has shown that tracking pain and finding patterns can give them a better understanding of how to manage their chronic pain condition~\cite{farrar2001clinical}.

When the body is injured, nerves send millions of signals to the brain, resulting in feelings of pain~\cite{wiech2008neurocognitive}. Since pain has a complex system, it is difficult to measure the pain signals. Researchers have investigated various methods to measure pain, such as using heart rate monitors to look for elevation in beat rate and correlating it to pain response~\cite{broucqsault2016measurement}.

However, these methods may not measure pain accurately since external factors such as anxiety, fear, and adrenaline can cause the same physical responses as pain~\cite{mcneil1992pain}.
The most common way to measure pain is for people to self-report their pain levels~\cite{ho1996review}.

\blfootnote{Poster presented at Graphics Interface Conference 2022\\
16-19 May - Montreal, QC, Canada\\
Copyright held by authors.\\ }

Clinical studies have also found that self-monitoring pain by chronic patients not only leads to patients having a greater understanding of how to cope with their condition, but they also become active contributors to their own care~\cite{blyth2007contribution}. Additionally, the collected data can help healthcare professionals support patients with managing their pain, leading to improved quality of care~\cite{farrar2001clinical}.
However, patients with chronic pain can often feel fatigued, which can make their everyday tasks more difficult and logging pain on top of their tasks can be demanding for them. Therefore, it is important to have an easy to use pain tracker for logging pain effortlessly~\cite{van2018tired}.

To be able to record pain accurately and swiftly, patients suffering from chronic diseases need to have a portable device to record their pain as they may feel a pain response while busy with daily activities. Wearable technology allows patients to keep pain tracking devices close at hand throughout the day. Additionally, there are mobile phone apps on the market where patients can record and track their pain levels~\cite{pfeifer2020mobile}.

Thus our goal in this research was to design, develop, and evaluate a portable pain tracking device for patients with chronic pain to record their pain levels. We proposed various portable device designs for tracking pain and received feedback from our healthcare professional collaborators on the designs. In an iterative process, we implemented our design ideas and developed PainBit. We evaluated the effectiveness of PainBit by asking patients to use the device for two weeks. Lastly, we conducted post study semi-structured interviews with patients to gain a better understanding of their experience using PainBit.

With this research, we contribute the design, development and evaluation  of a portable pain tracking device for chronic pain patients.

\begin{figure}[h]
    \centering
    \includegraphics[height=0.35\linewidth]{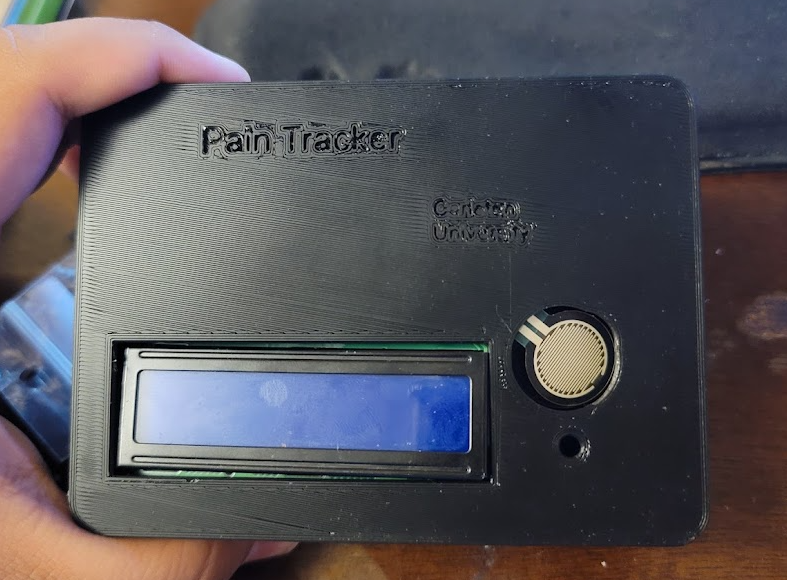}
    \caption{Pain tracker in a 3D printed case.}
    \label{fig:designOnecase}
\end{figure}

\section{Methods}

Upon receiving ethics approval, we conducted our study in two phases. In the first phase, we asked participants to collect their pain level whenever they experienced pain using our tracker for two weeks. In the second phase, we conducted semi-structured interviews with participants and asked them about their experience using the tracker.

\subsection{Patient Recruitment}
We recruited study participants through word of mouth and advertising the study on social media. Interested participants expressed their interest by contacting the researchers. We screened interested people who contacted us to match our inclusion criteria: adults (over the age of 18) who have been diagnosed with chronic pain (e.g., back pain, headache, Arthritis, cancer), have no cognitive disability (able to give informed consent), have hand motor abilities, and are able to speak English. Once we received the consent form, we sent PainBit to participants via post. We provided a \$20 gift card to participants in appreciation for their time.

\subsection{Patient Data Collection}

We asked participants to fill out a demographic form and collected information about their age, sex, chronic condition(s), and the number of years living with a chronic condition(s). We mailed the hardware tracker to participants and asked them to record their pain using the tracker for two weeks. The data was stored locally on the device. Once participants finished the two-week study period, we asked them to send the device back to us for data extraction.

\begin{figure}[h]
    \centering
    \includegraphics[height=0.45\linewidth]{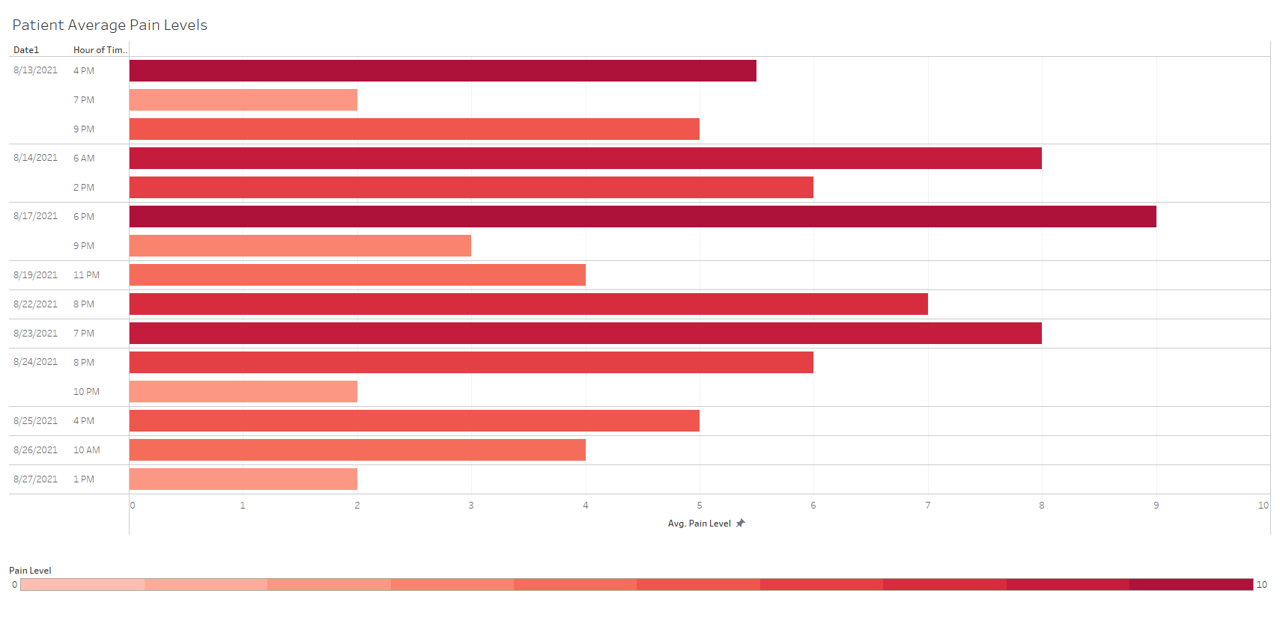}
    \caption{Fred (Patient 1) pain level data collected over 2 weeks.}
    \label{fig:pt1graph}
\end{figure}

\begin{figure}[h]
    \centering
    \includegraphics[height=0.45\linewidth]{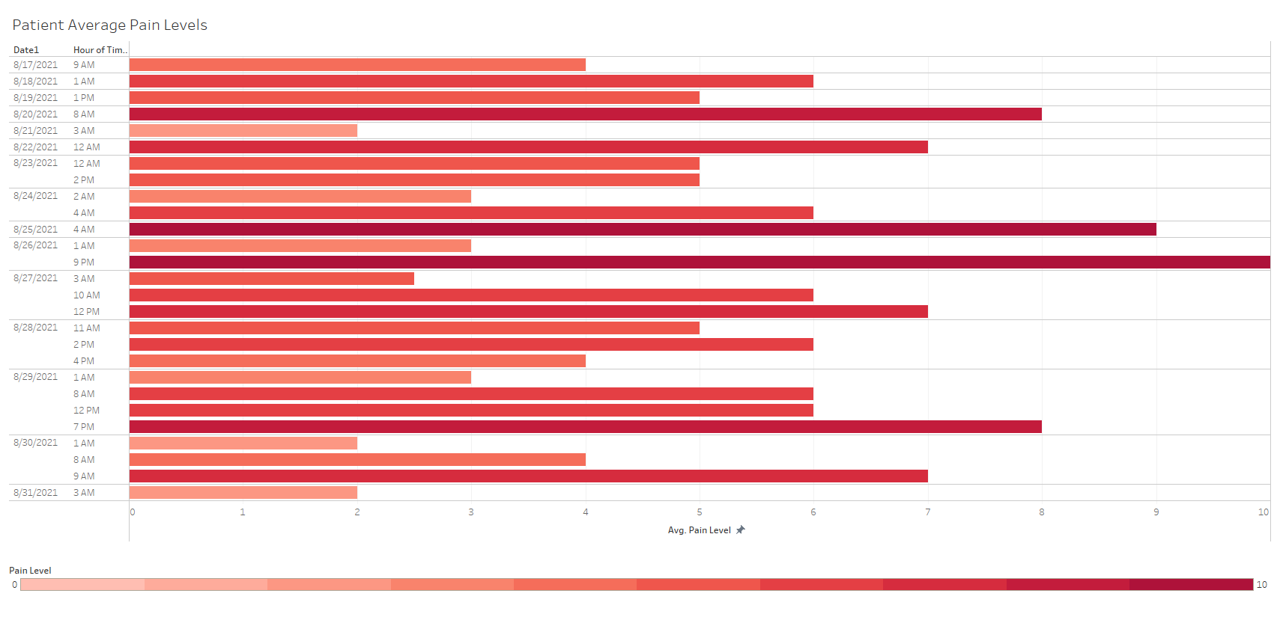}
    \caption{Dale (Patient 2) pain level data collected over 2 weeks}
    \label{fig:pt2graph}
\end{figure}

\subsection{Patient Interviews}
After the two-week data collection period, we conducted interviews with participants to learn about their experience using PainBit. Our team conducted the interview via Microsoft Teams. We audio-recorded and transcribed the interviews to analyze participants' inputs to improve our designs in future studies. Interviews lasted between 30 to 40 minutes.

We asked participants questions related to their experience with tracking their pain: How often did you use the device? Was there any day that you did not record your pain? If yes, why did you skip recording your pain? Did you use the device every time you felt pain? We also asked if participants had prior experience tracking their pain and What other device or app they used.
We also asked participants about their experience using the device and how accurately it displayed the pain they felt. Lastly, we asked if they had any suggestions to improve the device.
To inform the future design of an app to display the data, we asked participants how they would like to see the data collected from the tracker and what delivery format they would prefer to see their data.

Using applied thematic analysis~\cite{guest2011applied}, we qualitatively analyzed the interview responses by two analysts in our team. The analysts read the interview transcripts line by line and identified prominent codes and generated a codebook. The codes were then compared for agreement.  Then, analysts systematically grouped codes into candidate themes. Themes were compared and refined. Disagreements in the assignment of codes and themes were resolved by consensus arbitrated by the senior researcher in the team. 


\section{Results and Conclusion}

In collaboration with a healthcare provider, we designed a portable pain tracker device, PainBit, targeted for patients with chronic pain. Upon several iterations, we finalized PainBit implementation and conducted two case studies with patients using PainBit for two weeks and sharing their experience with us. 
From analyzing the results of our case studies, we identified several avenues to improve the design of PainBit. First, in the next iteration, we are planning to reduce the size and weight and design a wearable tracker that could be worn on a wrist. Second, to increase battery charging capacity, we plan on testing different batteries and assessing their weight and compatibility to increase battery life.
Lastly, we will investigate wearable designs with customization options. We are developing the second iteration of PainBit, which takes concerns such as battery life, size, weight, and form factor into account and improves in all those areas. The new form factor of the device will serve as wearable in the same fashion as a FitBit meant for pain tracking specifically.

The results of our study are a stepping stone for designing future targeted health tracking devices. We hope that in the long term, the results of these studies continue to contribute to the healthcare field to potentially offer enhanced care for patients with chronic pain.

\acknowledgments{
We would like to thank patients and healthcare
providers for the expert knowledge they brought to this
project. This work was supported in part by the NSERC Discovery grants.}

\bibliographystyle{abbrv-doi}

\bibliography{template}
\end{document}